\documentclass[epsfig,12pt]{article}
\usepackage{makeidx}
\usepackage{amsmath}
\usepackage{amsfonts}
\usepackage{amssymb}
\usepackage{graphicx}

\input epsf.sty
\newtheorem{theorem}{Theorem}
\newtheorem{acknowledgement}[theorem]{Acknowledgement}

\makeatletter \@addtoreset{equation}{section}
\renewcommand{\theequation}{\thesection.\arabic{equation}}
\input{tcilatex}
\begin{document}

\title{\vspace{-2cm}%
\rightline{\mbox{\small {LPHE-MS-15-08-rev}} \vspace
{1cm}} \textbf{Building SO}$_{10}$-\textbf{\ models with} $\mathbb{D}_{4}$
\textbf{symmetry}}
\author{R. Ahl Laamara$^{1,2}$, M. Miskaoui$^{1,2}$, E.H Saidi$^{2,3}$ \\
{\small 1. LPHE-Modeling and Simulations, Faculty Of Sciences, Rabat, Morocco%
}\\
{\small 2. Centre of Physics and Mathematics, CPM- Morocco}\\
{\small 3- International Centre for Theoretical Physics, Miramare, Trieste,
Italy}}
\maketitle

\begin{abstract}
Using characters of finite group representations and monodromy of matter
curves in F-GUT, we complete partial results in literature by building SO$%
_{10}$ models with dihedral $\mathbb{D}_{4}$ discrete symmetry. We first
revisit the $\mathbb{S}_{4}$-and $\mathbb{S}_{3}$-models from the discrete
group character view; then we extend the construction to $\mathbb{D}_{4}$.\
We find that there are three types of $SO_{10}\times \mathbb{D}_{4}$ models
depending on the ways the $\mathbb{S}_{4}$-triplets break down in terms of
irreducible $\mathbb{D}_{4}$- representations: $\left( \mathrm{\alpha }%
\right) $ as $\boldsymbol{1}_{_{+,-}}\oplus \boldsymbol{1}_{_{+,-}}\oplus
\boldsymbol{1}_{_{-,+}};$ or $\left( \mathrm{\beta }\right) \boldsymbol{\ 1}%
_{_{+,+}}\oplus \boldsymbol{1}_{_{+,-}}\oplus \boldsymbol{1}_{_{-,-}};$ or
also $\left( \mathrm{\gamma }\right) $ $\mathbf{1}_{_{+,-}}\oplus \mathbf{2}%
_{_{0,0}}$. Superpotentials and other features are also given.\newline
\end{abstract}


\section{Introduction}

In F-theory set-up, the study of GUT- models with discrete symmetries $%
\Gamma $ given by $\mathbb{S}_{n}$ permutation groups is generally done by
using the splitting spectral method \cite{1A}-\cite{7A}; see also \cite{8A}
and references therein. In the interesting case of SU$_{5}\times \Gamma $
models, the discrete $\Gamma $'s are mainly given by subgroups of the
permutation symmetry $\mathbb{S}_{5}$; in particular those subgroups like $%
\mathbb{S}_{4}$, $\mathbb{S}_{3}\times \mathbb{S}_{2},$ $\mathbb{S}_{3},%
\mathbb{S}_{2}\times \mathbb{S}_{2}$ and $\mathbb{S}_{2}\simeq \mathbb{Z}_{2}
$. To build SU$_{5}$ GUT- models with exotic discrete groups like the
alternating $\mathbb{A}_{4}$ and the dihedral $\mathbb{D}_{4}$, one needs
extra tools like Galois-theory \cite{1B,2B,3B,4B,5B,6B,8B}. In the case of SO%
$_{10}\times \Upsilon $ models the situation is quite similar except that
here the discrete $\Upsilon $'s are subgroups of the permutation symmetry $%
\mathbb{S}_{4}$. \newline
In \cite{1B}, an exhaustive study has been performed for several SO$%
_{10}\times \Upsilon $ models, broken down to SU$_{5}$, by using splitting
spectral method applied for the discrete subgroups $\Upsilon $ given by $%
\mathbb{S}_{3},$ $\mathbb{Z}_{3},$ $\mathbb{S}_{2}\times \mathbb{S}_{2}$ and
$\mathbb{S}_{2}$; the common denominator of these $\Upsilon $'s is that they
are subgroups of $\mathbb{S}_{4}$; the Weyl group of the $SU_{4}^{\perp }$
perpendicular symmetry to GUT gauge invariance in the E$_{8}$ breaking down
to $SO_{10}\times SU_{4}^{\perp }$. In this paper, we would like to complete
the analysis of \cite{1B} and subsequent studies by considering the case of
the order \emph{8} dihedral symmetry $\mathbb{D}_{4}$. This discrete group
is known to have two kinds of irreducible representations with dimensions 1
and 2; their multiplicities are read from the character relation $%
8=1^{2}+1^{2}+1^{2}+1^{2}+2^{2}$ teaching us the two following things: $%
\left( i\right) $ $\mathbb{D}_{4}$- group has four 1-$\dim $ representations
$\mathbf{1}_{p,q}$, including the trivial $\mathbf{1}_{+,+}$, the sign $%
\mathbf{1}_{-,+}$ as wells as two others $\mathbf{1}_{+,-}$ and $\mathbf{1}%
_{-,-}$; and $\left( ii\right) $ it has a unique 2-$\dim $ representation $%
\mathbf{2}_{0,0}$ with vanishing character vector $\left( 0,0\right) $.
These irreducible representations are phenomenologically interesting; first
because one of the four possible 1-$\dim $ representations of $\mathbb{D}_{4}
$ may be singled out to host the heaviest top- quark generation $\mathbf{16}%
_{3}$; and the unique 2-$\dim $ representation of $\mathbb{D}_{4}$ to
accommodate the two other $\mathbf{16}_{1,2}$ quark $u$- and $c$- families.
This picture goes with the idea of \textrm{\cite{5B}} where Yukawa matrix $%
\boldsymbol{Y}_{\left \{ c,s,t\right \} }$ for the $\left( u,c,t\right) $
quarks is approximated by a rank one matrix ($rank\boldsymbol{Y}=1$). There,
the $u,$ $c$ quarks are taken in the massless approximation; and the third
quark as a massive one. Moreover, $\mathbb{D}_{4}$- monodromy has also
enough different singlets to accommodate the three quark generations
independently; a property that may be used for studying superpotential
prototypes leading to higher rank mass matrix with hierarchical eigenvalues.%
\newline
The presentation is as follows: First we recall some useful aspects on $%
SO_{10}\times \Upsilon $ models concerning standard $\mathbb{S}_{4}$ and $%
\mathbb{S}_{3}$ discrete symmetries. By using irreducible representations $%
\boldsymbol{R}_{i}$ of these finite groups, we show how their character
functions $\mathrm{\chi }_{\boldsymbol{R}_{i}}$ can be used to characterise
the matter curve spectrum of these models. With these $\mathrm{\chi }_{%
\boldsymbol{R}_{i}}$ character tools at hand, we turn to build the above
mentioned three $SO_{10}\times \mathbb{D}_{4}$ models. We end this study by
a conclusion and discussions on building superpotentials.

\section{$SO_{10}\times \mathbb{S}_{4}$ model}

We begin by recalling that in $SO_{10}\times \mathbb{S}_{4}$ model of
F-theory GUTs, matter curves carry quantum numbers in the $SO_{10}\times
SU_{4}^{\bot }$ representations following from the breaking of the E$_{8}$
gauge symmetry of F-theory on elliptically fibered Calabi-Yau fourfold (CY4)
with 7-brane wrapping $\mathcal{S}_{GUT}$; the so called GUT surface \textrm{%
\cite{1A,1B,4B,5B,1C},\ }%
\begin{equation*}
\begin{tabular}{lll}
E & $\rightarrow $ & CY4 \\
&  & $\  \downarrow $ \\
&  & $\  \mathcal{B}_{3}$%
\end{tabular}%
\qquad ,\qquad \mathcal{B}_{3}\supset \mathcal{S}_{GUT}
\end{equation*}%
In this construction, the base $\mathcal{B}_{3}$ is a complex 3 $\dim $
manifold containing $\mathcal{S}_{GUT}$; and the fiber E is given by a
particular Tate representation of the complex elliptic curve, namely
\begin{equation}
y^{2}=x^{3}+b_{5}xyz+b_{4}x^{2}z+b_{3}yz^{2}+b_{2}xz^{3}+b_{0}z^{5}
\label{g}
\end{equation}%
where the homology classes $\left[ x\right] ,$ $\left[ y\right] ,$ $\left[ z%
\right] $, $\left[ b_{k}\right] ,$ associated with the holomorphic sections $%
x,$ $y,$ $z$ and $b_{k},$ are expressed in terms of the Chern class $%
c_{1}=c_{_{1}}\left( \mathcal{S}_{GUT}\right) $ of the tangent bundle of the
$\mathcal{S}_{GUT}$ surface; and the Chern class $-t$ of the normal bundle $%
\mathcal{N}_{\mathcal{S}_{GUT}|\mathcal{B}_{3}}$ as follows%
\begin{equation}
\begin{tabular}{lllllllll}
$\left[ y\right] $ & $=$ & $3\left( c_{1}-t\right) $ &  & $,$ &  & $\left[ z%
\right] $ & $=$ & $-t$ \\
$\left[ x\right] $ & $=$ & $2\left( c_{1}-t\right) $ &  & $,$ &  & $\left[
b_{k}\right] $ & $=$ & $\left( 6c_{1}-t\right) -kc_{1}$%
\end{tabular}%
\end{equation}%
\begin{equation*}
\end{equation*}%
Recall also that in $SO_{10}$ models building with discrete symmetries $%
\Gamma $, the $E_{8}$ symmetry of underlying F-theory on CY4 is broken down
to $SO_{10}\times \Gamma $; where $SO_{10}$ is the GUT gauge symmetry and $%
\Gamma $ a discrete monodromy group contained in $\mathbb{S}_{4}$. This
symmetric $\mathbb{S}_{4}$ is isomorphic to the Weyl symmetry group of the
perpendicular $SU_{4}^{\bot }$ to GUT symmetry inside $E_{8}$; that is the
commutant of $SO_{10}$ in the exceptional $E_{8}$ group of F-theory GUTs.%
\newline
From the decomposition of the $E_{8}$ adjoint representations down to $%
SO_{10}\times SU_{4}^{\bot }$ ones namely%
\begin{equation}
\begin{tabular}{lll}
$\mathbf{248}$ & $\rightarrow $ & $\left( \mathbf{45},\mathbf{1}_{\perp
}\right) \oplus \left( \mathbf{1},\mathbf{15}_{\perp }\right) \oplus $ \\
&  & $\left( \mathbf{16},\mathbf{4}_{\perp }\right) \oplus \left( \overline{%
\mathbf{16}},\mathbf{\bar{4}}_{\perp }\right) \oplus \left( \mathbf{10},%
\mathbf{6}_{\perp }\right) $%
\end{tabular}%
\end{equation}%
\begin{equation*}
\end{equation*}%
we learn the matter content of $SO_{10}\times SU_{4}^{\bot }$ theory; and
then of the desired curves spectrum of the $SO_{10}\times \mathbb{S}_{4}$
model. This spectrum is given by the following $SO_{10}$ multiplets, labeled
by four weights t$_{i}$ of the fundamental representation of $SU_{4}^{\bot
}, $%
\begin{equation}
\begin{tabular}{llll}
$\mathbf{16}_{t_{i}},$ & $\mathbf{16}_{-t_{i}},$ & $\mathbf{10}%
_{t_{i}+t_{j}},$ & $\mathbf{1}_{t_{i}-t_{j}}$%
\end{tabular}%
\end{equation}%
with the traceless condition%
\begin{equation}
t_{1}+t_{2}+t_{3}+t_{4}=0  \label{cd}
\end{equation}%
The discrete symmetry $\mathbb{S}_{4}$ acts by permutation of the t$_{i}$
curves; it leaves stable the constraint (\ref{cd}) as well as observable of
the model.\newline
In $SO_{10}\times \mathbb{S}_{4}$ theory, the 16-plets and the 10-plets are
thought of as \emph{reducible} multiplets of the $\mathbb{S}_{4}$ Weyl
symmetry of $SU_{4}^{\bot }$; however from the view of GUT phenomenology,
this $\mathbb{S}_{4}$ monodromy symmetry must be broken since it treats the
three GUT- generations on equal footing. But, for later use, we propose to
study first the structure of the $\mathbb{S}_{4}$ based model and some of
its basic properties; then turn back to study the breaking of $\mathbb{S}%
_{4} $ down to the subgroup $\mathbb{S}_{3}$; and after to the $\mathbb{D}%
_{4}$ we are interested in this paper.

\subsection{Spectrum in $\mathbb{S}_{4}$ model}

The $16_{t_{{\small i}}}$ components of the four 16-plets and the $10_{t_{%
{\small i}}+t_{j}}$ of the six 10-plets are related among themselves by $%
\mathbb{S}_{4}$ monodromy; they are respectively given by the zeros of the
holomorphic sections $b_{4}$ and $d_{6}$, describing the intersections of
the spectral covers $\mathcal{C}_{4}=0$ and $\mathcal{C}_{6}=0$ with the GUT
surface $s=0$. The defining eqs of these spectral covers are as shown on the
following table,
\begin{equation}
\begin{tabular}{|c|c|c|c|c|}
\hline
matters curves & weight & $\mathbb{S}_{4}$ & homology & holomorphic section
\\ \hline
$16_{t_{{\small i}}}$ & $t_{{\small i}}$ & $4$ & $\eta -4c_{1}$ & $%
b_{4}=b_{0}\dprod \limits_{i=1}^{4}t_{i}=0\left.
\begin{array}{c}
\text{ \ } \\
\text{ \  \ }%
\end{array}%
\right. $ \\ \hline
$10_{t_{{\small i}}+t_{j}}$ & $t_{{\small i}}+t_{j}$ & $6$ & $\eta ^{\prime
}-6c_{1}$ & $d_{6}=d_{0}\dprod \limits_{j>i=1}^{4}T_{ij}=0\left.
\begin{array}{c}
\text{ \ } \\
\text{ \  \ }%
\end{array}%
\right. $ \\ \hline
\end{tabular}
\label{1}
\end{equation}%
with \textrm{\cite{1B,4B,2C}}
\begin{equation}
\begin{tabular}{lllll}
$\mathcal{C}_{4}$ & $=$ & $\dsum \limits_{k=0}^{4}b_{k}s^{5-k}$ & $=$ & $%
b_{0}\dprod \limits_{i=1}^{4}\left( s-t_{i}\right) $ \\
&  &  &  &  \\
$\mathcal{C}_{6}$ & $=$ & $\dsum \limits_{k=0}^{6}d_{k}s^{6-k}$ & $=$ & $%
d_{0}\dprod \limits_{j>i=1}^{6}\left( s-T_{ij}\right) $%
\end{tabular}
\label{22}
\end{equation}%
and $T_{ij}\equiv t_{i}+t_{j}$. The homology class $\left[ b_{4}\right]
=\eta -4c_{1}$\ is obtained by using the relation $\left[ b_{0}\right] +\sum %
\left[ t_{i}\right] $ together with the canonical class $\left[ b_{0}\right]
=\eta $ and $\left[ t_{i}\right] =-c_{1};$ the same feature leads to $\left[
d_{6}\right] =\eta ^{\prime }-6c_{1}$.\newline
The $\mathbb{S}_{4}$ invariance of $\mathcal{C}_{4}$ is manifestly exhibited
\begin{equation}
\mathcal{C}_{4}=\frac{b_{0}}{4!}\sum_{\sigma \in \mathbb{S}_{4}}\dprod
\limits_{i=1}^{4}\left( s-t_{\sigma \left( i\right) }\right)
\end{equation}%
A similar relation is also valid for $\mathcal{C}_{6}$. In what follows, and
to fix ideas, we will think of the defining eqs of the spectral covers $%
\mathcal{C}_{4}$ and $\mathcal{C}_{6}$ as given by the last column of eqs(%
\ref{22}) involving $b_{0}$ and the four roots $t_{i}$ (resp $d_{0}$ and the
six $T_{ij}$'s).

\subsection{Characters in $SO_{10}\times \mathbb{S}_{4}$ model}

The discrete symmetry group of the $SO_{10}\times \mathbb{S}_{4}$ model has
\emph{24} elements that can be arranged into five conjugacy classes $%
\mathfrak{C}_{1},...,\mathfrak{C}_{5}$ with representatives given by
p-cycles $\left( 12...p\right) $ and products type $\mathrm{(\alpha \beta
)(\gamma \delta )}$ as shown on table (\ref{ca}). This finite group has five
irreducible representations $\boldsymbol{R}_{1},...,\boldsymbol{R}_{5}$
whose dimensions can be learnt from the usual relation $24=1^{2}+1^{2\prime
}+2^{2}+3^{2}+3^{2\prime }$ linking group order to the square of $\dim
\boldsymbol{R}_{i}$; their characters $\mathrm{\chi }_{\boldsymbol{R}_{j}}$,
describing mappings$:\mathbb{S}_{4}\rightarrow \mathbb{C}$, are conjugacy
class functions; associating to each class $\mathfrak{C}_{i}$ the numbers $%
\mathrm{\chi }_{ij}=\mathrm{\chi }_{\boldsymbol{R}_{j}}\left( \mathfrak{C}%
_{i}\right) $; whose explicit expressions are as given below%
\begin{equation}
\begin{tabular}{|l|l|l|l|l|l|l|}
\hline
$\mathfrak{C}_{i}$\TEXTsymbol{\backslash}irrep $\boldsymbol{R}_{j}$ & $\  \
\mathrm{\chi }_{_{\boldsymbol{I}}}$ & $\  \  \mathrm{\chi }_{_{\boldsymbol{3}%
^{\prime }}}$ & $\  \  \mathrm{\chi }_{_{\boldsymbol{2}}}$ & $\  \  \mathrm{\chi
}_{_{\boldsymbol{3}}}$ & $\  \  \mathrm{\chi }_{_{\epsilon }}$ & {\small order}
\\ \hline
$\mathfrak{C}_{1}\equiv \mathrm{\  \ e}$ & $\  \ 1\  \ $ & $\  \ 3\  \ $ & $\  \
2\  \ $ & $\  \ 3\  \ $ & $\  \ 1\  \ $ & $\  \ 1\  \ $ \\ \hline
$\mathfrak{C}_{2}\equiv \mathrm{(\alpha \beta )}$ & $\  \ 1$ & $-1$ & $\  \ 0$
& $\  \ 1$ & $-1$ & $\  \ 6$ \\ \hline
$\mathfrak{C}_{3}\equiv \mathrm{(\alpha \beta )(\gamma \delta )}$ & $\  \ 1$
& $-1$ & $\  \ 2$ & $-1$ & $\  \ 1$ & $\  \ 3$ \\ \hline
$\mathfrak{C}_{4}\equiv \mathrm{(\alpha \beta \gamma )}$ & $\  \ 1$ & $\  \ 0$
& $-1$ & $\  \ 0$ & $\  \ 1$ & $\  \ 8$ \\ \hline
$\mathfrak{C}_{5}\equiv \mathrm{(\alpha \beta \gamma \delta )}$ & $\  \ 1$ & $%
\  \ 1$ & $\  \ 0$ & $-1$ & $-1$ & $\  \ 6$ \\ \hline
\end{tabular}
\label{ca}
\end{equation}%
\begin{equation*}
\end{equation*}%
There are various manners to approach the properties of reducible $\mathcal{R%
}$ and irreducible $\boldsymbol{R}_{i}$ representations of the permutation
group $\mathbb{S}_{4}$; the natural way may be the one using graphic methods
based on the Young diagrams \textrm{\cite{1Y}}; where the five irreducible
representations are represented by \emph{5} diagrams as follows
\begin{equation}
1:\text{ \  \ }%
\begin{tabular}{|l|l|l|l|}
\hline
&  &  &  \\ \hline
\end{tabular}%
\qquad ,\qquad 2:\text{ \  \ }%
\begin{tabular}{|l|l|}
\hline
&  \\ \hline
&  \\ \hline
\end{tabular}%
\qquad ,\qquad 3:\text{ \  \ }%
\begin{tabular}{|l|ll}
\hline
&  & \multicolumn{1}{|l|}{} \\ \hline
&  &  \\ \cline{1-1}
\end{tabular}
\label{R1}
\end{equation}%
and%
\begin{equation}
3^{\prime }:\text{ \  \ }%
\begin{tabular}{|l|l}
\hline
& \multicolumn{1}{|l|}{} \\ \hline
&  \\ \cline{1-1}
&  \\ \cline{1-1}
\end{tabular}%
\qquad ,\qquad 1^{\prime }:\text{ \  \ }%
\begin{tabular}{|l|}
\hline
\\ \hline
\\ \hline
\\ \hline
\\ \hline
\end{tabular}
\label{R2}
\end{equation}%
But here we will deal with $\mathbb{S}_{4}$ by focussing on the properties
of its three generators $\left( a,b,c\right) $; these are basic operators of
the $\mathbb{S}_{4}$ group; and are generally chosen as given by the
transpositions $\tau _{1}=\left( 12\right) ,$ $\tau _{2}=\left( 23\right) $
and $\tau _{3}=\left( 34\right) $; they can be also taken as follows
\begin{equation}
\begin{tabular}{lll}
$A$ & $=$ & $\left( 12\right) =\tau _{1}$ \\
$B$ & $=$ & $\left( 123\right) =\tau _{1}\tau _{2}$ \\
$C$ & $=$ & $\left( 1234\right) =\tau _{1}\tau _{2}\tau _{3}$%
\end{tabular}%
\end{equation}%
These three generators are not independent seen that they are non
commutating operators, $AB\neq BA,$ $AC\neq CA$, $BC\neq CB$; a feature that
makes extracting total information a complicated matter; so we will restrict
to use their representation characters $\mathrm{\chi }_{\boldsymbol{R}%
_{i}}\left( A\right) $, $\mathrm{\chi }_{\boldsymbol{R}_{i}}\left( B\right) $
and $\mathrm{\chi }_{\boldsymbol{R}_{i}}\left( C\right) $ with $\boldsymbol{R%
}_{i}$ given by (\ref{R1}-\ref{R2}); we also use the sums of $\mathrm{\chi }%
_{\boldsymbol{R}_{i}}\left( G\right) $ and their products. To that purpose,
let us briefly recall some useful tools on $\mathbb{S}_{4}$- representations
that we illustrate on $SO_{10}\times \mathbb{S}_{4}$-theory. First notice
that the 4-$\dim $ permutation module $\mathcal{V}_{4}$ of the group $%
\mathbb{S}_{4}$ is interpreted in $SO_{10}\times \mathbb{S}_{4}$-model in
terms of the four matter curves $\mathbf{16}_{i}$; its 6-$\dim $
antisymmetric tensor product $\mathcal{V}_{6}=\left( \mathcal{V}_{4}\otimes
\mathcal{V}_{4}\right) _{antisy}$ as describing the Higgs curves $\mathbf{16}%
_{\left[ ij\right] }$; and the $\mathcal{V}_{4}\otimes \mathcal{V}_{4}^{\ast
}$ tensor product module is associated with flavons $\vartheta _{ij}$. All
these spaces are reducible under $\mathbb{S}_{4}$; and so permutation
operators of $\mathbb{S}_{4}$ can be generically decomposed as sums
\begin{equation}
n_{1}\mathbf{1}\oplus n_{1}^{\prime }\mathbf{1}^{\prime }\oplus n_{2}\mathbf{%
2}\oplus n_{3}^{\prime }\mathbf{3}\oplus n_{3}^{\prime }\mathbf{3}
\label{rp}
\end{equation}%
on the irreducible modules with some $n_{i}$ multiplicities; for example
\begin{equation}
\begin{tabular}{lll}
$\mathbf{4}$ & $\mathbf{=}$ & $\mathbf{1\oplus 3}$ \\
$\mathbf{6}$ & $=$ & $\mathbf{3\oplus 3}^{\prime }$%
\end{tabular}%
\qquad \leftrightarrow \qquad
\begin{tabular}{lll}
$\mathcal{V}_{4}$ & $\mathbf{=}$ & $\mathcal{V}_{1}\oplus \mathcal{V}_{3}$
\\
$\mathcal{V}_{6}$ & $=$ & $\mathcal{V}_{3}^{\prime }\oplus \mathcal{V}%
_{3}^{\prime \prime }$%
\end{tabular}
\label{v3}
\end{equation}%
with $\mathbf{6=}\left( \mathbf{4\otimes 4}\right) _{antisy}$. In practice,
the use of irreducible representations as in (\ref{rp}) for $SO_{10}\times
\mathbb{S}_{4}$ modeling is achieved by starting from eqs(\ref{1}); then
look for an adequate basis vector change of the weight $\left \{
t_{1},t_{2},t_{3},t_{4}\right \} $ into a new $\left \{
x_{0},x_{1},x_{2},x_{3}\right \} $ basis where one of the components, say $%
x_{0}$, has the form
\begin{equation}
x_{0}=\frac{1}{4}\left( t_{1}+t_{2}+t_{3}+t_{4}\right)  \label{x0}
\end{equation}%
this sum of weights is associated with the trivial representation $\mathbf{1}
$, the completely symmetric representation; it is invariant under $\mathbb{S}%
_{4}$; but also under all its subgroups including the $\mathbb{S}_{3}$ and
the $\mathbb{D}_{4}$ we will encounter below. The three other $\left(
x_{1},x_{2},x_{3}\right) =$ $\vec{x}$ transform as an irreducible triplet $%
\mathbf{3}$ under of $\mathbb{S}_{4}$; but differently under subgroups $%
\mathbb{S}_{3}$ and $\mathbb{D}_{4}$; their explicit expressions $\vec{x}=%
\vec{x}\left( t_{1},t_{2},t_{3},t_{4}\right) $ are given by;%
\begin{equation}
\begin{tabular}{lll}
$x_{1}$ & $=$ & $\frac{1}{4}\left( t_{1}+t_{2}-t_{3}-t_{4}\right) $ \\
$x_{2}$ & $=$ & $\frac{1}{4}\left( t_{1}-t_{2}+t_{3}-t_{4}\right) $ \\
$x_{3}$ & $=$ & $\frac{1}{4}\left( t_{1}-t_{2}-t_{3}+t_{4}\right) $%
\end{tabular}%
\end{equation}%
with (\ref{cd}) mapped to%
\begin{equation}
x_{0}+x_{1}+x_{2}+x_{3}=t_{1}
\end{equation}%
With the basis change of $\left \{ \left \vert t_{\mu }\right \rangle
\right
\} $ into $\left \{ \left \vert x_{0}\right \rangle ;\left \vert
x_{i}\right
\rangle \right \} $, the four matter $16_{t_{\mu }}$ and the
six Higgs $10_{t_{\mu }+t_{\nu }}$\ multiplets get splitted like%
\begin{equation}
16_{t_{\mu }}\rightarrow \left(
\begin{array}{c}
16_{0}^{\prime } \\
16_{i}^{\prime }%
\end{array}%
\right) _{\left \vert x\right \rangle }\qquad ,\qquad 10_{t_{\mu }+t_{\nu
}}\rightarrow \left(
\begin{array}{c}
10_{\left[ ij\right] }^{\prime } \\
10_{i}^{\prime }%
\end{array}%
\right) _{\left \vert x\otimes x\right \rangle }
\end{equation}%
In matrix notation with basis $\left \{ \left \vert x_{0}\right \rangle
;\left \vert x_{i}\right \rangle \right \} $, permutation operators $%
\boldsymbol{P}_{\sigma }^{\left( 4\right) }$ acting on $\mathcal{V}_{4}$ and
operators $\boldsymbol{P}_{\sigma }^{\left( 6\right) }$ on $\mathcal{V}_{6}$
have the representation
\begin{equation}
\boldsymbol{P}_{\sigma }^{\left( 4\right) }=\left(
\begin{array}{cc}
1 & 0_{1\times 3} \\
0_{3\times 1} & \mathcal{P}_{\sigma }^{\left( 3\right) }%
\end{array}%
\right) \qquad ,\qquad \boldsymbol{P}_{\sigma }^{\left( 6\right) }=\left(
\begin{array}{cc}
\mathcal{P}_{\sigma }^{\left( 3\right) } & 0_{3\times 3} \\
0_{3\times 3} & \mathcal{P}_{\sigma }^{\left( 3^{\prime }\right) }%
\end{array}%
\right)
\end{equation}%
To describe the new matter curves $\left \{ 16_{0}^{\prime },16_{i}^{\prime
}\right \} $ and $\left \{ 10_{i}^{\prime },10_{\left[ ij\right] }^{\prime
}\right \} $\ we shall use the character $\mathrm{\chi }_{\boldsymbol{R}%
_{i}} $ of the irreducible representations of $\mathbb{S}_{4}$ given by (\ref%
{ca}); in particular the character of the $\left( A,B,C\right) $ generators
of $\mathbb{S}_{4}$; and which we denote as $\mathrm{\chi }_{\boldsymbol{R}%
_{i}}^{\left( {\small G}\right) }$ where $G$ stands for $\left( A,B,C\right)
$,%
\begin{equation}
\begin{tabular}{|l|l|l|l|l|l|}
\hline
$\mathrm{\chi }_{\boldsymbol{R}_{j}}^{\left( G\right) }$ & $\mathrm{\chi }%
_{_{\boldsymbol{I}}}$ & $\mathrm{\chi }_{_{\boldsymbol{3}^{\prime }}}$ & $%
\mathrm{\chi }_{_{\boldsymbol{2}}}$ & $\mathrm{\chi }_{_{\boldsymbol{3}}}$ &
$\mathrm{\chi }_{_{\epsilon }}$ \\ \hline
$A$ & $1$ & $-1$ & $\  \ 0$ & $\  \ 1$ & $-1$ \\ \hline
$B$ & $1$ & $\  \ 0$ & $-1$ & $\  \ 0$ & $\  \ 1$ \\ \hline
$C$ & $1$ & $\  \ 1$ & $\  \ 0$ & $-1$ & $-1$ \\ \hline
\end{tabular}
\label{ac}
\end{equation}%
Because of a standard feature of the trace of direct sum of matrices namely $%
Tr\left( \mathcal{A}\oplus \mathcal{B}\right) =Tr\mathcal{A}+Tr\mathcal{B}$,
we also use the following property relating the characters of reducible $%
\mathcal{R}$ representations to their $\boldsymbol{R}_{i}$ irreducible
components%
\begin{equation}
\mathcal{R}=n_{1}\boldsymbol{R}_{1}\oplus n_{2}\boldsymbol{R}_{2}\qquad
\Rightarrow \qquad \mathrm{\chi }_{_{\mathcal{R}}}^{\left( {\small G}\right)
}=n_{1}\mathrm{\chi }_{_{\boldsymbol{R}_{1}}}^{\left( {\small G}\right)
}+n_{2}\mathrm{\chi }_{_{\boldsymbol{R}_{2}}}^{\left( {\small G}\right) }
\end{equation}%
For the example of the quartet $\mathbf{4=1\oplus 3}$ of the group $\mathbb{S%
}_{4}$, we have the relation $\mathrm{\chi }_{\mathbf{1\oplus 3}}^{\left(
{\small G}\right) }=\mathrm{\chi }_{\mathbf{1}}^{\left( {\small G}\right) }+%
\mathrm{\chi }_{\mathbf{3}}^{\left( {\small G}\right) }$; and remembering
the interpretation of the characters in terms of fix points of the
permutation symmetry; the character vector
\begin{equation}
\mathrm{\chi }_{\mathbf{4}}^{\left( {\small G}\right) }=\left( 2,1,0\right)
\equiv \mathrm{\chi }_{\mathbf{1}\oplus \mathbf{3}}^{\left( {\small G}%
\right) }  \label{23}
\end{equation}%
splits therefore as the sum of two terms: $\mathrm{\chi }_{\mathbf{1}%
}^{\left( {\small G}\right) }=\left( 1,1,1\right) $ and $\mathrm{\chi }_{%
\mathbf{3}}^{\left( {\small G}\right) }=\left( 1,0,-1\right) $; in agreement
with the character table (\ref{ac}). Applying also this property to $\mathbf{%
3\oplus 3}^{\prime }=\mathbf{6}$, we have $\mathrm{\chi }_{\mathbf{6}%
}^{\left( {\small G}\right) }=\mathrm{\chi }_{\mathbf{3}}^{\left( {\small G}%
\right) }+\mathrm{\chi }_{\mathbf{3}^{\prime }}^{\left( {\small G}\right) }$%
; from which we learn the value of the character of the reducible
6-dimensional representation of $\mathbb{S}_{4}$; it vanishes identically.
Therefore, the matter curves spectrum of the $SO_{10}\times \mathbb{S}_{4}$
model in the $\left \{ \left \vert x\right \rangle \right \} $ and $\left \{
\left \vert x\otimes x^{\prime }\right \rangle \right \} $ bases reads as
follows%
\begin{equation}
\begin{tabular}{|c|c|c|c|c|}
\hline
matters curves & $\mathbb{S}_{4}$ & homology & {\small U}${\small (1)}%
_{_{X}} ${\small \ }flux & character $\mathrm{\chi }_{_{\mathbf{R}}}^{\left(
{\small G}\right) }$ \\ \hline
$\left.
\begin{array}{c}
16_{0}^{\prime } \\
16_{i}^{\prime }%
\end{array}%
\right. $ & $\left.
\begin{array}{c}
\mathbf{1} \\
\mathbf{3}%
\end{array}%
\right. $ & $\left.
\begin{array}{c}
-c_{1} \\
\eta -3c_{1}%
\end{array}%
\right. $ & $\left.
\begin{array}{c}
0 \\
0%
\end{array}%
\right. $ & $\left.
\begin{array}{c}
\left( 1,1,1\right) \\
\left( 1,0,-1\right)%
\end{array}%
\right. $ \\ \hline
$\left.
\begin{array}{c}
10_{\left[ ij\right] }^{\prime } \\
10_{i}^{\prime }%
\end{array}%
\right. $ & $\left.
\begin{array}{c}
\mathbf{3}^{\prime } \\
\mathbf{3}%
\end{array}%
\right. $ & $\left.
\begin{array}{c}
-3c_{1} \\
\eta ^{\prime }-3c_{1}%
\end{array}%
\right. $ & $\left.
\begin{array}{c}
0 \\
0%
\end{array}%
\right. $ & $\left.
\begin{array}{c}
\left( -1,0,1\right) \\
\left( 1,0,-1\right)%
\end{array}%
\right. $ \\ \hline
\end{tabular}
\label{s4}
\end{equation}%
\begin{equation*}
\end{equation*}%
where the homology classes of the new matter curves $16_{0}^{\prime },$ $%
16_{i}^{\prime }$; and $10_{\left[ ij\right] }^{\prime }$, $10_{i}^{\prime }$
are derived from the homology of the reducible curve multiplets $16_{t_{\mu
}}$, $10_{t_{\mu }+t_{\nu }}$ of table eqs(\ref{1}) as follows%
\begin{equation}
\begin{tabular}{lll}
$\left[ 16_{t_{\mu }}\right] $ & $=$ & $\left[ 16_{0}^{\prime }\right] +%
\left[ 16_{i}^{\prime }\right] $ \\
$\left[ 10_{t_{\mu }+t_{\nu }}\right] $ & $=$ & $\left[ 10_{\left[ ij\right]
}^{\prime }\right] +\left[ 10_{i}^{\prime }\right] $%
\end{tabular}
\label{24}
\end{equation}
Notice that though the matter curves spectrum looks splitted, it is still
invariant under $\mathbb{S}_{4}$ monodromy. It is just a property of the $%
\left \vert x_{0,i}\right \rangle $ frame where the completely symmetric $%
x_{0} $ component weight, the centre of weights, is thought of as the origin
of the frame. Notice also that the $15$ flavons of the $SO_{10}\times
\mathbb{S}_{4}$ model split as $3\oplus 12;$ with the \emph{12} charged ones
splitting like $12=6\oplus 6^{\ast }$ where the $6$, currently denoted as $%
\vartheta _{+\left( t_{\mu }-t_{\nu }\right) }$ with $\mu <\nu $; and the
other $6^{\ast }$ with $\vartheta _{-\left( t_{\mu }-t_{\nu }\right) }$.
Because of the real values of the characters (\ref{ca}), the complex
adjoints $6^{\ast }=3^{\ast }\oplus 3^{\prime \ast }$ may be thought in
terms of dual representations namely $3^{\ast }\sim 3^{\prime }$ and $%
3^{\prime \ast }\sim 3$; so the characters for the \emph{12} charged flavons
read as
\begin{equation}
\begin{tabular}{|c|c|}
\hline
flavons & character $\mathrm{\chi }_{_{\mathbf{R}}}^{\left( {\small G}%
\right) }$ \\ \hline
$\left.
\begin{array}{c}
1_{i} \\
1_{\left[ ij\right] }%
\end{array}%
\right. $ & $\left.
\begin{array}{c}
\left( 1,0,-1\right) \\
\left( -1,0,1\right)%
\end{array}%
\right. $ \\ \hline
$\left.
\begin{array}{c}
1_{\left[ ij\right] } \\
1_{i}%
\end{array}%
\right. $ & $\left.
\begin{array}{c}
\left( -1,0,1\right) \\
\left( 1,0,-1\right)%
\end{array}%
\right. $ \\ \hline
\end{tabular}%
\end{equation}

\section{Building $SO_{10}\times \mathbb{D}_{4}$ models}

First we describe the key idea of our method that we illustrate on the
example of $SO_{10}\times \mathbb{S}_{3}$ model; seen that $\mathbb{S}_{3}$
is a subgroup of $\mathbb{S}_{4}$ just as $\mathbb{D}_{4}$. Then we turn to
study the $SO_{10}\times \mathbb{D}_{4}$ theory; and derive its matter
curves spectrum and their characters; comments on the superpotentials $W$ of
$\mathbb{D}_{4}$ models and others aspects will be given in conclusion and
discussion section.

\subsection{Revisiting $SO_{10}\times \mathbb{S}_{3}$ model}

To engineer $SO_{10}\times \Gamma $ models with discrete symmetries $\Gamma $
contained in $\mathbb{S}_{4}$, we have to break the $\mathbb{S}_{4}$
symmetry down to its subgroup $\Gamma $. Seen that $\mathbb{S}_{4}$ has
\emph{30} subgroups; one ends with a proletariat of $SO_{10}$ models with
discrete symmetries; some of these monodromies are related amongst others by
similarity transformations. For example, $\mathbb{S}_{4}$ has four $\mathbb{S%
}_{3}$ subgroups obtained by fixing one of the four $t_{i}$ roots of the
spectral cover $\mathcal{C}_{4}$ of eq(\ref{22}); but these $\mathbb{S}_{3}$
groups are isomorphic to each other; and so it is enough to consider just
one of them; say the one fixing the weight $t_{4}$; and permuting the other
three $t_{1},$ $t_{2},$ $t_{3}$; i.e:%
\begin{equation}
\begin{tabular}{lll}
$\  \  \  \  \  \  \  \  \  \  \  \  \sigma \left( t_{4}\right) $ & $=$ & $t_{4}$ \\
$\sigma \left( \left \{ t_{1},t_{2},t_{3}\right \} \right) $ & $=$ & $\left
\{ t_{1},t_{2},t_{3}\right \} $%
\end{tabular}%
\end{equation}%
leading to $\sigma \in \mathbb{S}_{4}/\mathcal{J}$, with $\mathcal{J}%
=\left
\langle t_{4}-\sigma \left( t_{4}\right) \right \rangle $; it is
isomorphic to $\mathbb{S}_{3}$. Let us describe rapidly our method of
engineering $SO_{10}\times \mathbb{S}_{3}$ model; and extend later this
construction to the case of the order \emph{8} dihedral $\mathbb{D}_{4}$.
\newline
Starting from the matter spectrum of the $SO_{10}\times \mathbb{S}_{4}$
model (\ref{1}), we can derive the properties of the matter curves of the $%
SO_{10}\times \mathbb{S}_{3}$ model by using the breaking pattern%
\begin{equation}
\mathbb{S}_{4}\qquad \rightarrow \qquad \mathbb{S}_{3}\times \mathbb{S}_{1}
\end{equation}%
where $\mathbb{S}_{1}$ factor is associated with the fixed weight $t_{4}$;
it may be interpreted in terms of a conserved U$_{1}^{\perp }$ symmetry
inside SU$_{4}^{\perp }$. The descent from $\mathbb{S}_{4}$ to $\mathbb{S}%
_{3}$ model is known to be due turning on an abelian flux piercing the $%
\mathbb{S}_{4}$ - matter and Higgs multiplets; and is commonly realised by
the splitting spectral method like $\mathcal{C}_{4}=\mathcal{C}_{3}\times
\mathcal{C}_{1}$ and $\mathcal{C}_{6}=\mathcal{\tilde{C}}_{3}\times \mathcal{%
\tilde{C}}_{3}^{\prime }$.\  \newline
\textrm{By using the gauge 2-form field strength }$\boldsymbol{F}_{X}$ of
the U$\left( 1\right) _{X}$ gauge symmetry considered in \textrm{\cite{1B}};
and by thinking of the $\mathbb{S}_{4}$ invariance of the $SO_{10}\times
\mathbb{S}_{4}$ model of table (\ref{s4}) in terms of vanishing quantized
flux $\mathcal{F}_{\xi }^{X}$ of the 2-form $\boldsymbol{F}_{X}$ over a
2-cycle $\xi $ in the homology of base of the CY4, namely
\begin{equation}
\begin{tabular}{lll}
$\mathcal{F}_{\xi }^{X}|_{H}$ & $=$ & $\dint \nolimits_{\xi }\boldsymbol{F}%
_{X}=0$ \\
$\mathcal{F}_{c_{1}}^{X}|_{H}$ & $=$ & $\dint \nolimits_{c_{1}}\boldsymbol{F}%
_{X}=0$ \\
$\mathcal{F}_{\eta }^{X}|_{H}$ & $=$ & $\dint \nolimits_{\eta }\boldsymbol{F}%
_{X}=0$%
\end{tabular}%
\qquad ,\qquad H=\mathbb{S}_{4};
\end{equation}%
then the breaking of $\mathbb{S}_{4}$ monodromy down to discrete subgroups $%
H $ may be realised\footnote{%
\ In table 1 of ref \textrm{\cite{1B}, }the SO$_{10}\times \mathbb{S}_{4}$%
\textrm{\ }model monodromy is broken down to\textrm{\ }$\mathbb{Z}_{2}$%
\textrm{.}} as in table 1 of ref \textrm{\cite{1B}} by giving non zero value
to $\mathcal{F}_{\xi }^{X}$ as follows
\begin{equation}
\begin{tabular}{llll}
$\mathcal{F}_{\xi }^{X}|_{H}$ & $=$ & $N$ & $\neq 0$ \\
$\mathcal{F}_{c_{1}}^{X}|_{H}$ & $=$ & $0$ &  \\
$\mathcal{F}_{\eta }^{X}|_{H}$ & $=$ & $0$ &
\end{tabular}%
\qquad ,\qquad H\subset \mathbb{S}_{4}  \label{fn}
\end{equation}%
where $N$ is an integer. The above relation is in fact just an equivalent
statement of breaking monodromies by using the splitting spectral cover
method where covers $\mathcal{C}_{n}$ are factorised as product $\mathcal{C}%
_{n_{1}}\times \mathcal{C}_{n_{2}}$ with $n_{1}+n_{2}=n$. The extra
relations $\mathcal{F}_{c_{1}}^{X}|_{H}=\mathcal{F}_{\eta }^{X}|_{H}=0$ are
the usual conditions to avoid Green- Schwarz mass for the U$\left( 1\right)
_{X}$ gauge field potential \textrm{\cite{2A,3A,1B}}. \newline
In our approach, the effect of the abelian non zero flux $\mathcal{F}_{\xi
}^{X}|_{H}$ is interpreted in terms of piercing the irreducible $\mathbb{S}%
_{4}$- triplets $16_{i}$, $10_{i}$ and $10_{\left[ ij\right] }$ as follows
\begin{eqnarray}
&&\left.
\begin{array}{c}
\text{ } \\
\mathbf{16}_{i}%
\end{array}%
\right. \qquad \left.
\begin{array}{c}
\text{ } \\
\rightarrow%
\end{array}%
\right. \qquad
\begin{tabular}{l|ll}
${\small curves}$ & \multicolumn{2}{|l}{${\small number}$} \\ \hline
$\mathbf{16}_{a}$ & $3-N$ & $=2$ \\
$\mathbf{16}_{3}$ & $3-2N$ & $=1$%
\end{tabular}
\\
&&\text{ }  \notag \\
&&\text{ }  \notag \\
&&\left.
\begin{array}{c}
\text{ } \\
\mathbf{10}_{i}%
\end{array}%
\right. \qquad \left.
\begin{array}{c}
\text{ } \\
\rightarrow%
\end{array}%
\right. \qquad
\begin{tabular}{l|l}
${\small curves}$ & ${\small number}$ \\ \hline
$\mathbf{10}_{a}$ & $2$ \\
$\mathbf{10}_{3}$ & $1$%
\end{tabular}
\\
&&\text{ }  \notag \\
&&\text{ }  \notag \\
&&\left.
\begin{array}{c}
\text{ } \\
\mathbf{10}_{\left[ ij\right] }%
\end{array}%
\right. \qquad \left.
\begin{array}{c}
\text{ } \\
\rightarrow%
\end{array}%
\right. \qquad
\begin{tabular}{l|l}
${\small curves}$ & ${\small number}$ \\ \hline
$\mathbf{10}_{\left[ a3\right] }$ & $2$ \\
$\mathbf{10}_{\left[ 12\right] }$ & $1$%
\end{tabular}%
\end{eqnarray}%
In geometrical words, flux piercing of the $16_{i}$ triplet corresponds to
breaking the isotropy of the 3-dimension subspace $\mathcal{V}_{3}$ of eq(%
\ref{v3}) as a direct sum $\mathcal{V}_{1}\oplus \mathcal{V}_{2}$; the same
breaking happens for the spaces $\mathcal{V}_{3}^{\prime }$ and $\mathcal{V}%
_{3}^{\prime \prime }$ associated with $10_{i}$ and $10_{\left[ ij\right] };$
and to any non trivial representation of $\mathbb{S}_{4}$. Due to the non
zero flux, the $\mathbb{S}_{4}$- triplet $\left \{
x_{1},x_{2},x_{3}\right
\} $ gets splitted into a $\mathbb{S}_{3}$- doublet
$\left \{ \left \vert y_{1}\right \rangle ,\left \vert y_{2}\right \rangle
\right \} $ and a $\mathbb{S}_{3}$- singlet $\left \{ \left \vert
y_{3}\right \rangle \right \} $; the new $y_{\mu }$ weights are related to
the previous weights x$_{\mu }$ as follows
\begin{equation}
\begin{tabular}{lll}
$y_{1}$ & $=$ & $\frac{1}{3}\left( x_{1}+x_{2}-2x_{3}\right) $ \\
$y_{2}$ & $=$ & $\frac{1}{3}\left( x_{1}-2x_{2}+x_{3}\right) $%
\end{tabular}%
\end{equation}%
and%
\begin{equation}
\begin{tabular}{lll}
$y_{3}$ & $=$ & $\frac{1}{3}\left( x_{1}+x_{2}+x_{3}\right) $ \\
$y_{0}$ & $=$ & $x_{0}$%
\end{tabular}
\label{0x}
\end{equation}%
their expressions in terms of the canonical $t_{\mu }$- weights are obtained
through the relations $x_{0}=x_{0}\left( t_{\mu }\right) ,$ $%
x_{i}=x_{i}\left( t_{\mu }\right) $. In the $\left \{ \left \vert
y_{i}\right \rangle \right \} $ basis, the distribution of the abelian flux
among the new directions in the 4-$\dim $ permutation module $\mathcal{V}%
_{4} $ is as illustrated on the following $4\times 4$ traceless matrix%
\begin{equation}
\left(
\begin{array}{cccc}
4N &  &  &  \\
& -N &  &  \\
&  & -N &  \\
&  &  & -2N%
\end{array}%
\right) _{\left \vert y_{i}\right \rangle }
\end{equation}%
Recall that the permutation group $\mathbb{S}_{3}$ has order 6, three
conjugacy classes and three irreducible representations; $6=1^{2}+1^{\prime
2}+2^{2}$; its character table is as follows%
\begin{equation}
\begin{tabular}{|l|l|l|l|l|}
\hline
$\mathfrak{C}_{i}$\TEXTsymbol{\backslash}irrep $\boldsymbol{R}_{j}$ & $%
\mathrm{\chi }_{_{\boldsymbol{I}}}$ & $\mathrm{\chi }_{_{\boldsymbol{2}}}$ &
$\mathrm{\chi }_{_{\epsilon }}$ & {\small order} \\ \hline
$\mathfrak{C}_{1}\equiv \mathrm{\  \ e}$ & $\  \ 1$ \  & $\  \ 2$ \  & $\  \ 1$
\  & $\  \ 1$ \\ \hline
$\mathfrak{C}_{2}\equiv \mathrm{(\alpha \beta )}$ & $\  \ 1$ & $\ 0$ & $-1$ &
$\  \ 3$ \\ \hline
$\mathfrak{C}_{3}\equiv \mathrm{(\alpha \beta \gamma )}$ & $\  \ 1$ & $-1$ & $%
\  \ 1$ & $\  \ 2$ \\ \hline
\end{tabular}
\label{ch}
\end{equation}%
\begin{equation*}
\end{equation*}%
The\ group $\mathbb{S}_{3}$ has two non commuting generators that can be
chosen like $A=\left( 12\right) $ and $B=\left( 123\right) $ with characters
as in above table. Using similar notations as for the case of the group $%
\mathbb{S}_{4}$; in particular the property regarding the relation between
the characters of reducible $\mathcal{R}$ and irreducible $\boldsymbol{R}%
_{i} $ representations of $\mathbb{S}_{3}$, we have
\begin{equation}
\mathrm{\chi }_{_{\mathcal{R}}}^{\left( {\small g}\right) }=n_{1}\mathrm{%
\chi }_{_{\boldsymbol{R}_{1}}}^{\left( {\small g}\right) }+n_{2}\mathrm{\chi
}_{_{\boldsymbol{R}_{2}}}^{\left( {\small g}\right) }
\end{equation}%
where now $g=\left( A,B\right) $. Moreover, by using the interpretation of
the character $\mathrm{\chi }_{_{\mathcal{R}}}^{\left( {\small g}\right) }$
in terms of fix points of $\mathbb{S}_{3}$- permutations; we have, on one
hand $\mathrm{\chi }_{\mathbf{1\oplus 2}}^{\left( {\small g}\right) }=\left(
1,0\right) $; and on the other hand $\mathrm{\chi }_{\mathbf{1}}^{\left(
{\small g}\right) }=\left( 1,1\right) $ and $\mathrm{\chi }_{\mathbf{2}%
}^{\left( {\small g}\right) }=\left( 0,-1\right) ,$ in agreement with
character table (\ref{ch}),%
\begin{equation}
\left( 1,0\right) =\left( 1,1\right) +\left( 0,-1\right)
\end{equation}%
Furthermore, by remembering that the $SO_{10}\times \mathbb{S}_{3}$-
modeling is done by starting from $\mathbb{S}_{4}$- multiplets and, due to
non zero flux, they decompose into irreducible $\mathbb{S}_{3}$
representations. For the matter sector of the $SO_{10}\times \mathbb{S}_{3}$
model, the four 16-plets transforming in quartet multiplet $\mathbf{4}$
decomposes in terms of irreducible $\mathbb{S}_{3}$ representations as the
sum of two singlets $\mathbf{1}_{1}$, $\mathbf{1}_{2}$ and a doublet as
follows%
\begin{equation}
\mathbf{4}=\mathbf{1}_{1}\oplus \mathbf{1}_{2}\oplus \mathbf{2}
\end{equation}%
However, seen that in $\mathbb{S}_{3}$ representation theory, we have two
kinds of singlets namely the trivial $e$ and the sign $\epsilon $, we must
determine the nature of the singlets $\mathbf{1}_{1}$ and $\mathbf{1}_{2}$
involved in above equation. While one of these singlets; say the $\mathbf{1}%
_{1}$ should be a trivial singlet as it corresponds just to the weight $%
y_{0}=x_{0}$ of eqs(\ref{x0}-\ref{0x}), we still have to determine the
nature of the $\mathbf{1}_{2}$; but this is also a trivial object since it
can be checked directly from our explicit construction as it corresponds
precisely to the weight y$_{3}$ given by eq(\ref{0x}). Nevertheless, this
result can be also obtained by using the following consistency relation
given by the restriction of the $\mathbb{S}_{4}$ relation $\mathrm{\chi }_{%
\mathbf{4}}^{\left( {\small G}\right) }=\mathrm{\chi }_{\mathbf{1}}^{\left(
{\small G}\right) }+\mathrm{\chi }_{\mathbf{3}}^{\left( {\small G}\right) }$
down to its subgroup $\mathbb{S}_{3}$; namely
\begin{equation}
\left. \mathrm{\chi }_{\mathbf{4}}^{\left( {\small G}\right) }\right \vert _{%
\mathbb{S}_{3}}=\left. \mathrm{\chi }_{\mathbf{1}}^{\left( {\small G}\right)
}\right \vert _{\mathbb{S}_{3}}+\left. \mathrm{\chi }_{\mathbf{3}}^{\left(
{\small G}\right) }\right \vert _{\mathbb{S}_{3}}  \label{sd}
\end{equation}%
By restricting this relationship to the $g=\left( A,B\right) $ generators of
$\mathbb{S}_{3}$ and using $\mathrm{\chi }_{\mathbf{3}}^{\left( {\small g}%
\right) }=\mathrm{\chi }_{\mathbf{1}\oplus \mathbf{2}}^{\left( {\small g}%
\right) }$, the restricted eq(\ref{sd}) reads as $\mathrm{\chi }_{\mathbf{1}%
}^{\left( {\small g}\right) }+\mathrm{\chi }_{\mathbf{1}\oplus \mathbf{2}%
}^{\left( {\small g}\right) }$ and leads to
\begin{equation}
\left( 2,1\right) =\left( 1,1\right) +\left( 0,-1\right) +\left( k,l\right)
\end{equation}%
from which we learn that $\left( k,l\right) $ should be equal to $\left(
1,1\right) $; and so the $\mathbf{1}_{2}$ singlet must be in the trivial
representation. A similar reasoning leads to the decompositions of the $%
\mathbb{S}_{4}$- characters $\mathrm{\chi }_{\mathbf{3}}^{\left( {\small G}%
\right) }$ and $\mathrm{\chi }_{\mathbf{3}^{\prime }}^{\left( {\small G}%
\right) }$ down to their $\mathbb{S}_{3}$- counterparts; we find $\mathrm{%
\chi }_{\mathbf{3}}^{\left( {\small g}\right) }=\mathrm{\chi }_{\mathbf{1}%
\oplus \mathbf{2}}^{\left( {\small g}\right) }$ and $\mathrm{\chi }_{\mathbf{%
3}^{\prime }}^{\left( {\small g}\right) }=\mathrm{\chi }_{\mathbf{1}^{\prime
}\mathbf{\oplus 2}}^{\left( {\small g}\right) }$ with the value%
\begin{equation}
\begin{tabular}{lll}
$\mathrm{\chi }_{\mathbf{1\oplus 2}}^{\left( {\small g}\right) }$ & $=$ & $%
\left( 1,1\right) +\left( 0,-1\right) $ \\
$\mathrm{\chi }_{\mathbf{1}^{\prime }\mathbf{\oplus 2}}^{\left( {\small g}%
\right) }$ & $=$ & $\left( -1,1\right) +\left( 0,-1\right) $%
\end{tabular}%
\end{equation}%
Therefore, the $\mathbb{S}_{3}$- matter curves spectrum following from the
breaking of (\ref{s4}) is given by%
\begin{equation}
\begin{tabular}{|c|c|c|c|c|}
\hline
matters curves & $\mathbb{S}_{3}$ & homology & {\small U}${\small (1)}%
_{_{X}} ${\small \ }flux & character $\mathrm{\chi }_{_{\boldsymbol{R}%
}}^{\left( {\small g}\right) }$ \\ \hline
$\left.
\begin{array}{c}
16_{0}^{\prime \prime } \\
16_{i}^{\prime \prime } \\
16_{3}^{\prime \prime }%
\end{array}%
\right. $ & $\left.
\begin{array}{c}
\mathbf{1}_{+,+} \\
\mathbf{2}_{0,-} \\
\mathbf{1}_{+,+}%
\end{array}%
\right. $ & $\left.
\begin{array}{c}
4\xi -c_{1} \\
\eta -2c_{1}-2\xi \\
-c_{1}-2\xi%
\end{array}%
\right. $ & $\left.
\begin{array}{c}
4N \\
-2N \\
-2N%
\end{array}%
\right. $ & $\left.
\begin{array}{c}
\left( 1,1\right) \\
\left( 0,-1\right) \\
\left( 1,1\right)%
\end{array}%
\right. $ \\ \hline
$\left.
\begin{array}{c}
10_{\left[ 12\right] }^{\prime \prime } \\
10_{\left[ i3\right] }^{\prime \prime } \\
10_{i}^{\prime \prime } \\
10_{3}^{\prime \prime }%
\end{array}%
\right. $ & $\left.
\begin{array}{c}
\mathbf{1}_{-,+} \\
\mathbf{2}_{0,-} \\
\mathbf{2}_{0,-} \\
\mathbf{1}_{+,+}%
\end{array}%
\right. $ & $\left.
\begin{array}{c}
2\xi -c_{1} \\
-2c_{1}-2\xi \\
\eta ^{\prime }-2c_{1}-2\xi \\
2\xi -c_{1}%
\end{array}%
\right. $ & $\left.
\begin{array}{c}
2N \\
-2N \\
-2N \\
2N%
\end{array}%
\right. $ & $\left.
\begin{array}{c}
\left( -1,1\right) \\
\left( 0,-1\right) \\
\left( 0,-1\right) \\
\left( 1,1\right)%
\end{array}%
\right. $ \\ \hline
\end{tabular}
\label{s3}
\end{equation}%
\begin{equation*}
\end{equation*}%
where we have indexed the irreducible $\mathbb{S}_{3}$- representations by
their characters. In this table, $\xi $ is a new 2-cycle; and the integer $N$
standing for the the abelian flux $\int_{\xi }\boldsymbol{F}_{X}$ inducing
the breaking down to $\mathbb{S}_{3}$; see also eq(\ref{fn}). Moreover,
using properties on characters of tensor products; in particular the typical
relations $X_{p,q}\otimes Y_{k,l}=\sum Z_{n,m}$ requiring $pk=\sum n$ and $%
ql=\sum m$, we obtain the following $\mathbb{S}_{3}$- algebra \
\begin{equation}
\begin{tabular}{lll}
$\mathbf{2}_{0,-}\otimes \mathbf{2}_{0,-}$ & $=$ & $\mathbf{1}_{+,+}\oplus
\mathbf{1}_{-,+}\oplus \mathbf{2}_{0,-}$ \\
$\mathbf{1}_{p,+}\otimes \mathbf{2}_{0,-}$ & $=$ & $\mathbf{2}_{0,-}$ \\
$\mathbf{1}_{p,+}\otimes \mathbf{1}_{+,+}$ & $=$ & $\mathbf{1}_{p,+}$ \\
$\mathbf{1}_{+,+}\otimes \mathbf{1}_{-,+}$ & $=$ & $\mathbf{1}_{-,+}$%
\end{tabular}%
\end{equation}%
with $p=\pm 1.$

\subsection{$SO_{10}\times \mathbb{D}_{4}$ models}

We first recall useful aspects on the dihedral group $\mathbb{D}_{4}$ and
the characters of their representations; then we build three $SO_{10}\times
\mathbb{D}_{4}$- models and give their matter and Higgs curves spectrum by
using specific properties of the breaking of $\mathbb{S}_{4}$ down to $%
\mathbb{D}_{4}$.

\subsubsection{Characters in $\mathbb{D}_{4}$ models}

The dihedral $\mathbb{D}_{4}$ group is an order \emph{8} subgroup of $%
\mathbb{S}_{4}$ with the usual decomposition property $%
8=1_{1}+1_{2}+1_{3}+1_{4}+2^{2}$ showing that, generally speaking, $\mathbb{D%
}_{4}$ has \emph{5} irreducible representations: four of them are 1-
dimensional, denoted as $\mathbf{1}_{i}$; the fifth has 2-dimensions. The
finite group $\mathbb{D}_{4}$ has two generators $a$ and $c$ satisfying the
relations $a^{2}=1$, $c^{4}=1,$ and $aca=c^{3}$ with $c^{3}=c^{-1}$ and $%
a=a^{-1}$. It has \emph{5} conjugation classes $\mathfrak{C}_{\alpha }$
given by%
\begin{equation}
\begin{tabular}{lllll}
$\mathfrak{C}_{1}\equiv \left \{ e\right \} $ & , & $\mathfrak{C}_{2}\equiv
\left \{ c^{2}\right \} $ & , & $\mathfrak{C}_{3}\equiv \left \{
c,c^{3}\right \} $ \\
$\mathfrak{C}_{4}\equiv \left \{ a,c^{2}a\right \} $ & $,$ & $\mathfrak{C}%
_{5}\equiv \left \{ ca,c^{3}a\right \} $ &  &
\end{tabular}%
\end{equation}%
The character table of the irreducible representation of $\mathbb{D}_{4}$ is
as follows%
\begin{equation}
\begin{tabular}{|l|l|l|l|l|l|l|}
\hline
$\mathfrak{C}_{i}$\TEXTsymbol{\backslash}$\mathrm{\chi }_{\boldsymbol{R}%
_{j}} $ & $\  \  \mathrm{\chi }_{_{\mathbf{1}_{1}}}$ & $\  \  \mathrm{\chi }_{_{%
\mathbf{1}_{2}}}$ & $\  \  \mathrm{\chi }_{_{\mathbf{1}_{3}}}$ & $\  \  \mathrm{%
\chi }_{\mathbf{1}_{4}}$ & $\  \  \mathrm{\chi }_{_{2}}$ & number \\ \hline
$\mathfrak{C}_{1}$ & $\  \ 1\  \ $ & $\  \ 1\  \ $ & $\  \ 1\  \ $ & $\  \ 1\  \ $ &
$\  \ 2\  \ $ & $\  \ 1\  \ $ \\ \hline
$\mathfrak{C}_{2}$ & $\  \ 1$ & $\  \ 1$ & $\  \ 1$ & $\  \ 1$ & $-2$ & $\  \ 1$
\\ \hline
$\mathfrak{C}_{3}$ & $\  \ 1$ & $\  \ 1$ & $-1$ & $-1$ & $\  \ 0$ & $\  \ 2$ \\
\hline
$\mathfrak{C}_{4}$ & $\  \ 1$ & $-1$ & $\  \ 1$ & $-1$ & $\  \ 0$ & $\  \ 2$ \\
\hline
$\mathfrak{C}_{5}$ & $\  \ 1$ & $-1$ & $-1$ & $\  \ 1$ & $-0$ & $\  \ 2$ \\
\hline
\end{tabular}
\label{hc}
\end{equation}%
To make contact between the three $\left( A,B,C\right) \equiv G$\ generators
of $\mathbb{S}_{4}$ and the two above $\left( a,c\right) $ generators of $%
\mathbb{D}_{4}$; notice that the dihedral group has no irreducible 3-cycles;
then we have%
\begin{equation}
a=\left. A\right \vert _{\mathbb{D}_{4}}\qquad c=\left. C\right \vert _{%
\mathbb{D}_{4}}  \label{pr}
\end{equation}%
and therefore the following
\begin{equation}
\begin{tabular}{|l|l|l|l|l|l|}
\hline
$\mathfrak{\chi }_{ij}^{\left( g\right) }$ & $\  \  \mathrm{\chi }_{_{\mathbf{1%
}_{1}}}$ & $\  \  \mathrm{\chi }_{_{\mathbf{1}_{2}}}$ & $\  \  \mathrm{\chi }_{_{%
\mathbf{1}_{3}}}$ & $\  \  \mathrm{\chi }_{\mathbf{1}_{4}}$ & $\  \  \mathrm{%
\chi }_{_{2}}$ \\ \hline
$a$ & $\  \ 1$ & $-1$ & $\  \ 1$ & $-1$ & $\  \ 0$ \\ \hline
$c$ & $\  \ 1$ & $\  \ 1$ & $-1$ & $-1$ & $\  \ 0$ \\ \hline
\end{tabular}%
\end{equation}%
exhibiting explicitly the difference between the four singlets. For
convenience, we denote now on the \emph{5} irreducible representations of
the dihedral group by their characters as follows
\begin{equation}
\begin{tabular}{lllllllll}
$\mathbf{1}_{+,+}$ & , & $\mathbf{1}_{+,-}$ & , & $\mathbf{1}_{-,+}$ & , & $%
\mathbf{1}_{-,-}$ & , & $\mathbf{2}_{0,0}$%
\end{tabular}%
\end{equation}%
Notice that from the above character table for the $\left( a,c\right) $
generators of $\mathbb{D}_{4}$, we can build three different kinds of $%
\mathbb{D}_{4}$ algebras; these are:

\begin{itemize}
\item $\mathbb{D}_{4}$\emph{- algebra I}
\begin{equation}
\begin{tabular}{lll}
$\mathbf{2}_{0,0}\otimes \mathbf{2}_{0,0}$ & $=$ & $\mathbf{1}_{+,+}\oplus
\mathbf{1}_{-,-}\oplus \mathbf{1}_{+,-}\oplus \mathbf{1}_{-,+}$ \\
$\mathbf{1}_{p,q}\otimes \mathbf{1}_{p,q}$ & $=$ & $\mathbf{1}_{pp^{\prime
},qq^{\prime }}$ \\
$\mathbf{1}_{p,q}\otimes \mathbf{2}_{0,0}$ & $=$ & $\mathbf{2}_{0,0}$%
\end{tabular}%
\end{equation}

\item $\mathbb{D}_{4}$\emph{- algebra II}
\begin{equation}
\begin{tabular}{lll}
$\mathbf{2}_{0,0}\otimes \mathbf{2}_{0,0}$ & $=$ & $\mathbf{1}_{+,+}\oplus
\mathbf{1}_{-,-}\oplus \mathbf{2}_{0,0}$ \\
$\mathbf{1}_{p,q}\otimes \mathbf{1}_{p,q}$ & $=$ & $\mathbf{1}_{pp^{\prime
},qq^{\prime }}$ \\
$\mathbf{1}_{p,q}\otimes \mathbf{2}_{0,0}$ & $=$ & $\mathbf{2}_{0,0}$%
\end{tabular}%
\end{equation}

\item $\mathbb{D}_{4}$\emph{- algebra III}%
\begin{equation}
\begin{tabular}{lll}
$\mathbf{2}_{0,0}\otimes \mathbf{2}_{0,0}$ & $=$ & $\mathbf{1}_{+,-}\oplus
\mathbf{1}_{-,+}\oplus \mathbf{2}_{0,0}$ \\
$\mathbf{1}_{p,q}\otimes \mathbf{1}_{p,q}$ & $=$ & $\mathbf{1}_{pp^{\prime
},qq^{\prime }}$ \\
$\mathbf{1}_{p,q}\otimes \mathbf{2}_{0,0}$ & $=$ & $\mathbf{2}_{0,0}$%
\end{tabular}%
\end{equation}
\end{itemize}

\  \  \  \  \  \  \  \newline
To engineer the $SO_{10}\times \mathbb{D}_{4}$ models, we proceed as in the
case of $\mathbb{S}_{3}$; we start from the spectrum (\ref{s4}) of the $%
\mathbb{S}_{4}$ model and break its monodromy down to $\mathbb{D}_{4}$ by
turning on an appropriate abelian flux (\ref{fn}) that pierces the $1+3$
matter curves $16_{0}\oplus 16_{i}$ and the $3+3^{\prime }$ Higgs curves $%
10_{i}\oplus 10_{\left[ ij\right] }$. We distinguish three cases depending
the way the $\mathbb{S}_{4}$- triplet is splitted; they are described below:

\subsubsection{Three $SO_{10}\times \mathbb{D}_{4}$ models}

First, we study the two cases where the $\mathbb{S}_{4}$- triplet $16_{i}$
and the two $\mathbb{S}_{4}$- triplets $10_{i}\oplus 10_{\left[ ij\right] }$
are completely reduced down in terms of $\mathbb{D}_{4}$ singlets. Then, we
study the other possible case where these triplets are decomposed as sums of
a $\mathbb{D}_{4}$- singlet and a $\mathbb{D}_{4}$- doublet.

\  \  \  \  \

$\alpha )$ $SO_{10}\times \mathbb{D}_{4}$\emph{- models I and II}\newline
We will see below that according to the values of the characters of the $%
\mathbb{D}_{4}$- generators, there are two configurations for the splitting $%
3=1+1+1$; they depend on presence or absence of trivial $1_{++}$
representation as given below:

\begin{itemize}
\item a splitting with no $\mathbb{D}_{4}$- trivial singlet%
\begin{equation}
\mathbf{3=1}_{+,-}\oplus \mathbf{1}_{+,-}\oplus \mathbf{1}_{-,+}
\end{equation}

\item a splitting with a $\mathbb{D}_{4}$- trivial singlet%
\begin{equation}
\mathbf{3=1}_{+,+}\oplus \mathbf{1}_{-,-}\oplus \mathbf{1}_{+,-}
\end{equation}
\end{itemize}

\  \  \  \  \newline
To establish this claim, we start from the character of the generators of $%
\mathbb{S}_{4}$; and consider the computation of the following restriction
down to $\mathbb{D}_{4}$ subgroup%
\begin{equation}
\left. \mathrm{\chi }_{\mathbf{4}}^{\left( {\small g}\right) }\right \vert _{%
\mathbb{D}_{4}}=\left. \mathrm{\chi }_{\mathbf{1}}^{\left( {\small g}\right)
}\right \vert _{\mathbb{D}_{4}}+\left. \mathrm{\chi }_{\mathbf{3}}^{\left(
{\small g}\right) }\right \vert _{\mathbb{D}_{4}}  \label{13}
\end{equation}%
Thinking of the triplet $\mathbf{3}$ as the direct sum of three singlets $%
\mathbf{1}_{1}\oplus \mathbf{1}_{2}\oplus \mathbf{1}_{3}$, we then have the
following character relationship
\begin{equation}
\left. \mathrm{\chi }_{\mathbf{4}}^{\left( {\small g}\right) }\right \vert _{%
\mathbb{D}_{4}}=\left. \mathrm{\chi }_{\mathbf{1}_{1}}^{\left( {\small g}%
\right) }\right \vert _{\mathbb{D}_{4}}+\left. \mathrm{\chi }_{\mathbf{1}%
_{2}}^{\left( {\small g}\right) }\right \vert _{\mathbb{D}_{4}}+\left.
\mathrm{\chi }_{\mathbf{1}_{3}}^{\left( {\small g}\right) }\right \vert _{%
\mathbb{D}_{4}}+\left. \mathrm{\chi }_{\mathbf{1}_{4}}^{\left( {\small g}%
\right) }\right \vert _{\mathbb{D}_{4}}
\end{equation}%
By combining eqs(\ref{23}) and (\ref{pr}), we learn that the left hand side
of equation is nothing but $\left. \mathrm{\chi }_{\mathbf{4}}^{\left(
{\small g}\right) }\right \vert _{\mathbb{D}_{4}}=\left( 2,0\right) $;
while, by using the characters of the singlets of table (\ref{hc}), the
right hand side decomposes like%
\begin{equation}
\left( 2,0\right) =\left( 1,1\right) +\left( k_{1},l_{1}\right) +\left(
k_{2},l_{2}\right) +\left( k_{3},l_{3}\right)
\end{equation}%
with $k_{i}=\pm 1$ and $l_{i}=\pm 1$. Explicitly%
\begin{equation}
\begin{tabular}{lll}
$k_{1}+k_{2}+k_{3}$ & $=$ & $+1$ \\
$l_{1}+l_{2}+l_{3}$ & $=$ & $-1$%
\end{tabular}%
\end{equation}%
which can be solved in two manners: $\left( \mathbf{i}\right) $ either as
\begin{equation}
\left( 2,0\right) =\left( 1,1\right) +\left( 1,-1\right) +\left( 1,-1\right)
+\left( -1,1\right)  \label{a}
\end{equation}%
involving three kinds of irreducibles 1-$\dim $ representations of $\mathbb{D%
}_{4}$; the trivial representation with character $\left( 1,1\right) $ and
two others with characters $\left( -1,1\right) $ appearing once and the $%
\left( 1,-1\right) $ appearing twice; or $\left( \mathbf{ii}\right) $ like%
\begin{equation}
\left( 2,0\right) =\left( 1,1\right) +\left( 1,-1\right) +\left( 1,1\right)
+\left( -1,-1\right)  \label{b}
\end{equation}%
where the trivial representation appears twice. Accordingly, we have the
following curves spectrums:

\  \  \  \  \  \  \  \  \  \  \

$\bullet $ $SO_{10}\times \mathbb{D}_{4}$- \emph{model I}\newline
It is given by the decomposition (\ref{a}); its matter spectrum reads as
\begin{equation}
\begin{tabular}{|c|c|c|c|c|}
\hline
matters curves & $\mathbb{D}_{4}$ & homology & {\small U}${\small (1)}%
_{_{X}} ${\small \ }flux & character $\mathrm{\chi }_{_{\boldsymbol{R}%
}}^{\left( {\small g}\right) }$ \\ \hline
$\left.
\begin{array}{c}
\mathbf{16}_{0}^{\prime \prime } \\
\mathbf{16}_{1}^{\prime \prime } \\
\mathbf{16}_{2}^{\prime \prime } \\
\mathbf{16}_{3}^{\prime \prime }%
\end{array}%
\right. $ & $\left.
\begin{array}{c}
\mathbf{1}_{+,+} \\
\mathbf{1}_{+,-} \\
\mathbf{1}_{+,-} \\
\mathbf{1}_{-,+}%
\end{array}%
\right. $ & $\left.
\begin{array}{c}
4\xi -c_{1} \\
\eta -c_{1}-\xi \\
-c_{1}-\xi \\
-c_{1}-2\xi%
\end{array}%
\right. $ & $\left.
\begin{array}{c}
4N \\
-N \\
-N \\
-2N%
\end{array}%
\right. $ & $\left.
\begin{array}{c}
\left( 1,1\right) \\
\left( 1,-1\right) \\
\left( 1,-1\right) \\
\left( -1,1\right)%
\end{array}%
\right. $ \\ \hline
$\left.
\begin{array}{c}
\mathbf{10}_{\left[ 12\right] }^{\prime \prime } \\
\mathbf{10}_{\left[ 13\right] }^{\prime \prime } \\
\mathbf{10}_{\left[ 23\right] }^{\prime \prime } \\
\mathbf{10}_{1}^{\prime \prime } \\
\mathbf{10}_{2}^{\prime \prime } \\
\mathbf{10}_{3}^{\prime \prime }%
\end{array}%
\right. $ & $\left.
\begin{array}{c}
\mathbf{1}_{-,+} \\
\mathbf{1}_{+,-} \\
\mathbf{1}_{+,-} \\
\mathbf{1}_{+,-} \\
\mathbf{1}_{-,+} \\
\mathbf{1}_{-,+}%
\end{array}%
\right. $ & $\left.
\begin{array}{c}
2\xi -c_{1} \\
-c_{1}-\xi \\
-c_{1}-\xi \\
\eta ^{\prime }-2c_{1}-\xi \\
-c_{1}-\xi \\
2\xi -c_{1}%
\end{array}%
\right. $ & $\left.
\begin{array}{c}
2N \\
-N \\
-N \\
-N \\
-N \\
2N%
\end{array}%
\right. $ & $\left.
\begin{array}{c}
\left( -1,1\right) \\
\left( 1,-1\right) \\
\left( 1,-1\right) \\
\left( 1,-1\right) \\
\left( -1,1\right) \\
\left( -1,1\right)%
\end{array}%
\right. $ \\ \hline
\end{tabular}
\label{d4}
\end{equation}%
\begin{equation*}
\end{equation*}%
and%
\begin{equation}
\begin{tabular}{|c|c|}
\hline
flavons & character $\mathrm{\chi }_{_{\mathbf{R}}}^{\left( {\small g}%
\right) }$ \\ \hline
$\left.
\begin{array}{c}
\mathbf{1}_{-,+} \\
\mathbf{1}_{+,-} \\
\mathbf{1}_{+,-} \\
\mathbf{1}_{+,-} \\
\mathbf{1}_{-,+} \\
\mathbf{1}_{-,+}%
\end{array}%
\right. $ & $\left.
\begin{array}{c}
\left( -1,1\right) \\
\left( 1,-1\right) \\
\left( 1,-1\right) \\
\left( 1,-1\right) \\
\left( -1,1\right) \\
\left( -1,1\right)%
\end{array}%
\right. $ \\ \hline
\end{tabular}%
\end{equation}%
together with their adjoints.%
\begin{equation*}
\end{equation*}

$\bullet $ $SO_{10}\times \mathbb{D}_{4}$- \emph{model II}\newline
It is given by the decomposition (\ref{b}) with matter curve spectrum like%
\begin{equation}
\begin{tabular}{|c|c|c|c|c|}
\hline
matters curves & $\mathbb{D}_{4}$ & homology & {\small U}${\small (1)}%
_{_{X}} ${\small \ }flux & character $\mathrm{\chi }_{_{\boldsymbol{R}%
}}^{\left( {\small g}\right) }$ \\ \hline
$\left.
\begin{array}{c}
\mathbf{16}_{0}^{\prime \prime } \\
\mathbf{16}_{1}^{\prime \prime } \\
\mathbf{16}_{2}^{\prime \prime } \\
\mathbf{16}_{3}^{\prime \prime }%
\end{array}%
\right. $ & $\left.
\begin{array}{c}
\mathbf{1}_{+,+} \\
\mathbf{1}_{+,-} \\
\mathbf{1}_{-,-} \\
\mathbf{1}_{+,+}%
\end{array}%
\right. $ & $\left.
\begin{array}{c}
4\xi -c_{1} \\
\eta -c_{1}-\xi \\
-c_{1}-\xi \\
-c_{1}-2\xi%
\end{array}%
\right. $ & \multicolumn{1}{|c|}{$\left.
\begin{array}{c}
4N \\
-N \\
-N \\
-2N%
\end{array}%
\right. $} & $\left.
\begin{array}{c}
\left( 1,1\right) \\
\left( 1,-1\right) \\
\left( -1,-1\right) \\
\left( 1,1\right)%
\end{array}%
\right. $ \\ \hline
$\left.
\begin{array}{c}
\mathbf{10}_{\left[ 12\right] }^{\prime \prime } \\
\mathbf{10}_{\left[ 13\right] }^{\prime \prime } \\
\mathbf{10}_{\left[ 23\right] }^{\prime \prime } \\
\mathbf{10}_{1}^{\prime \prime } \\
\mathbf{10}_{2}^{\prime \prime } \\
\mathbf{10}_{3}^{\prime \prime }%
\end{array}%
\right. $ & $\left.
\begin{array}{c}
\mathbf{1}_{+,-} \\
\mathbf{1}_{+,+} \\
\mathbf{1}_{-,-} \\
\mathbf{1}_{-,+} \\
\mathbf{1}_{-,-} \\
\mathbf{1}_{+,+}%
\end{array}%
\right. $ & $\left.
\begin{array}{c}
2\xi -c_{1} \\
-c_{1}-\xi \\
-c_{1}-\xi \\
\eta ^{\prime }-2c_{1}-\xi \\
-c_{1}-\xi \\
2\xi -c_{1}%
\end{array}%
\right. $ & \multicolumn{1}{|c|}{$\left.
\begin{array}{c}
2N \\
-N \\
-N \\
-N \\
-N \\
2N%
\end{array}%
\right. $} & $\left.
\begin{array}{c}
\left( 1,-1\right) \\
\left( 1,1\right) \\
\left( -1,-1\right) \\
\left( -1,1\right) \\
\left( -1,-1\right) \\
\left( 1,1\right)%
\end{array}%
\right. $ \\ \hline
\end{tabular}
\label{d5}
\end{equation}%
and%
\begin{equation*}
\end{equation*}%
\begin{equation}
\begin{tabular}{|c|c|}
\hline
flavons & character $\mathrm{\chi }_{_{\mathbf{R}}}^{\left( {\small g}%
\right) }$ \\ \hline
$\left.
\begin{array}{c}
\mathbf{1}_{+,-} \\
\mathbf{1}_{+,+} \\
\mathbf{1}_{-,-} \\
\mathbf{1}_{-,+} \\
\mathbf{1}_{-,-} \\
\mathbf{1}_{+,+}%
\end{array}%
\right. $ & $\left.
\begin{array}{c}
\left( 1,-1\right) \\
\left( 1,1\right) \\
\left( -1,-1\right) \\
\left( -1,1\right) \\
\left( -1,-1\right) \\
\left( 1,1\right)%
\end{array}%
\right. $ \\ \hline
\end{tabular}%
\end{equation}

$\beta )$ $SO_{10}\times \mathbb{D}_{4}$\emph{- model III}\newline
This model corresponds to the splitting $\mathbf{3}=\mathbf{1}\oplus \mathbf{%
2}$; the restricted character relation (\ref{13}) decomposes like
\begin{equation}
\left. \mathrm{\chi }_{\mathbf{4}}^{\left( {\small g}\right) }\right \vert _{%
\mathbb{D}_{4}}=\mathrm{\chi }_{\mathbf{1}}^{\left( {\small g}\right) }+%
\mathrm{\chi }_{\mathbf{1}^{\prime }}^{\left( {\small g}\right) }+\mathrm{%
\chi }_{\mathbf{2}}^{\left( {\small g}\right) }
\end{equation}%
leading to%
\begin{equation}
\left( 2,0\right) =\left( 1,1\right) +\left( k_{1},l_{1}\right) +\left(
k_{2},l_{2}\right)
\end{equation}%
and then $\left( k_{1},l_{1}\right) +\left( k_{2},l_{2}\right) =\left(
1,-1\right) $. However seen that the character $\mathrm{\chi }_{\mathbf{2}%
}^{\left( {\small g}\right) }=\left( 0,0\right) $; it follows that%
\begin{equation}
\boldsymbol{3}=\mathbf{2}_{0,0}\oplus \mathbf{1}_{+,-}
\end{equation}%
Therefore the matter spectrum of the $SO_{10}\times \mathbb{D}_{4}$- model
III is given by
\begin{equation}
\begin{tabular}{|c|c|c|c|c|}
\hline
matters curves & $\mathbb{D}_{4}$ & homology & {\small U}${\small (1)}%
_{_{X}} ${\small \ }flux & character $\mathrm{\chi }_{_{\boldsymbol{R}%
}}^{\left( {\small g}\right) }$ \\ \hline
$\left.
\begin{array}{c}
\mathbf{16}_{0}^{\prime \prime } \\
\mathbf{16}_{i}^{\prime \prime } \\
\mathbf{16}_{3}^{\prime \prime }%
\end{array}%
\right. $ & $\left.
\begin{array}{c}
\mathbf{1}_{+,+} \\
\mathbf{2}_{0,0} \\
\mathbf{1}_{+,-}%
\end{array}%
\right. $ & $\left.
\begin{array}{c}
4\xi -c_{1} \\
\eta -2c_{1}-2\xi \\
-c_{1}-2\xi%
\end{array}%
\right. $ & $\left.
\begin{array}{c}
4N \\
-2N \\
-2N%
\end{array}%
\right. $ & $\left.
\begin{array}{c}
\left( 1,1\right) \\
\left( 0,0\right) \\
\left( 1,-1\right)%
\end{array}%
\right. $ \\ \hline
$\left.
\begin{array}{c}
\mathbf{10}_{\left[ 12\right] }^{\prime \prime } \\
\mathbf{10}_{\left[ i3\right] }^{\prime \prime } \\
\mathbf{10}_{i}^{\prime \prime } \\
\mathbf{10}_{3}^{\prime \prime }%
\end{array}%
\right. $ & $\left.
\begin{array}{c}
\mathbf{1}_{-,+} \\
\mathbf{2}_{0,0} \\
\mathbf{2}_{0,0} \\
\mathbf{1}_{+,-}%
\end{array}%
\right. $ & $\left.
\begin{array}{c}
2\xi -c_{1} \\
-2c_{1}-2\xi \\
\eta ^{\prime }-2c_{1}-2\xi \\
2\xi -c_{1}%
\end{array}%
\right. $ & $\left.
\begin{array}{c}
2N \\
-2N \\
-2N \\
2N%
\end{array}%
\right. $ & $\left.
\begin{array}{c}
\left( -1,1\right) \\
\left( 0,0\right) \\
\left( 0,0\right) \\
\left( 1,-1\right)%
\end{array}%
\right. $ \\ \hline
\end{tabular}
\label{d6}
\end{equation}%
\begin{equation*}
\end{equation*}%
and%
\begin{equation}
\begin{tabular}{|c|c|}
\hline
flavons & character $\mathrm{\chi }_{_{\mathbf{R}}}^{\left( {\small g}%
\right) }$ \\ \hline
$\left.
\begin{array}{c}
\mathbf{1}_{-,+} \\
\mathbf{2}_{0,0} \\
\mathbf{2}_{0,0} \\
\mathbf{1}_{+,-}%
\end{array}%
\right. $ & $\left.
\begin{array}{c}
\left( -1,1\right) \\
\left( 0,0\right) \\
\left( 0,0\right) \\
\left( 1,-1\right)%
\end{array}%
\right. $ \\ \hline
\end{tabular}
\label{42}
\end{equation}%
\begin{equation*}
\end{equation*}%
Notice that in tables (\ref{d4}-\ref{d6}), the homology classes of the new
curves are obtained as usual by requiring the sum of their homology classes
to be equal to the class of their mother matter curve in $\mathbb{S}_{4}$
model.\ The extra 2-cycle class $\xi $ and the corresponding integral flux $%
N=\int_{\xi }\boldsymbol{F}_{X}$ are associated with the breaking of $%
\mathbb{S}_{4}$ down to $\mathbb{D}_{4}$; the 2-form $\boldsymbol{F}_{X}$ is
as in eq(\ref{fn}).

\section{Conclusion and discussions}

In this work, we have used characters of discrete group representations to
approach the engineering of $SO_{10}\times \Gamma $ models with discrete
monodromy $\Gamma $ contained in $\mathbb{S}_{4}$; this method generalises
straightforwardly to SU$_{5}\times \Gamma $ and its breaking down to MSSM.
In this construction, curves of the GUT- models are described by the
characters $\mathrm{\chi }_{\boldsymbol{R}_{i}}$ of the irreducible
representations $\boldsymbol{R}_{i}$ of the discrete group $\Gamma $.\newline
After having introduced the idea of the character based method; we have
revisited the study of the $\mathbb{S}_{4}$- and $\mathbb{S}_{3}$- models
from the view of monodromy irreducible representations and their characters;
see table (\ref{s4}) for the case $\Gamma =\mathbb{S}_{4}$, the table (\ref%
{s3}) for $\mathbb{S}_{3}$- model. Then, we have extended the construction
to the dihedral group where we have found that there are three $%
SO_{10}\times \mathbb{D}_{4}$- models with curves spectrum as in tables (\ref%
{d4}), (\ref{d5}) and (\ref{d6}). \newline
The approach introduced and developed in this study has two remarkable
particularities: $\left( i\right) $ first it allows to build GUT- models
with subgroups inside the $\mathbb{S}_{4}$ permutation symmetry like $%
\mathbb{D}_{4}$ and $\mathbb{A}_{4}$ without resorting to the use of Galois
theory. The latter is known to lead to non linear constraints on the
holomorphic sections of the spectral covers; and requires more involved
tools for their solutions. $\left( ii\right) $ second, it is based on a
natural quantity of discrete symmetry groups; namely the characters of their
representations; known as basic objects to deal with finite order symmetry
groups.

\  \

\emph{Building superpotentials using characters}\newline
The character based method developed in this study can be also used to build
superpotentials. For the case of $SO_{10}\times \mathbb{D}_{4}$-models;
superpotentials $W$ invariant under discrete symmetry $\mathbb{D}_{4}$ are
obtained by requiring their character as $\mathrm{\chi }_{_{\boldsymbol{R}%
}}^{\left( {\small g}\right) }\left( W\right) =\left( 1,1\right) $. By
focussing on the models (\ref{d4}-\ref{d5}); and denoting the matter curves
with $\left( p,q\right) $ character as $\mathbf{16}_{p,q}$ and the Higgs
curve with character $\left( \alpha ,\beta \right) $ like $\mathbf{10}%
_{\alpha ,\beta }$; and restricting to typical tri-coupling superpotential,
we have the following candidate%
\begin{equation*}
W=\sum \lambda _{pq}^{p^{\prime }q^{\prime }}\mathbf{16}_{p,q}\otimes
\mathbf{16}_{p^{\prime },q^{\prime }}\otimes \mathbf{10}_{1/pp^{\prime
},1/qq^{\prime }}
\end{equation*}%
\ More general expressions can be written down by implementing flavons as
well. However, to write down phenomenologically acceptable Yukawa couplings
that agree with low energy effective field constraints such as reproducing a
MSSM like spectrum and suppressing rapid proton decay, one must have a
diagonal tree-level Yukawa coupling for the heaviest top-quark family; one
also need to introduce extra discrete symmetries such as R-parity or the $%
\mathbb{Z}_{2}$ geometric symmetry of \textrm{\cite{1X,1B,4B}} to rule out
undesired couplings.\ Diagonal tree level 3-couplings for the top-quark
family have been studied in \textrm{\cite{1B}} for the case of $\mathbb{S}%
_{3}$ and its subgroups; they extend directly to our analysis; it reads in
character language as
\begin{equation*}
W_{tree}=\lambda _{{\small top}}\text{ }\mathbf{16}_{+-}\otimes \mathbf{16}%
_{+-}\otimes \mathbf{10}_{++}
\end{equation*}
singling out the spectrum of the $\mathbb{D}_{4}$- model (\ref{d5}). For the
spectrum (\ref{d6}), a superpotential with a diagonal Yukawa coupling for a
top-quark in a $\mathbb{D}_{4}$-singlet requires flavons; if taking the
top-quark matter curve in $\mathbf{16}_{+-}$ as above; a superpotential $%
W_{\ast }$ candidate would have the form
\begin{equation*}
W_{\ast }\sim \mathbf{16}_{+-}\otimes \mathbf{16}_{+-}\otimes \mathbf{10}%
_{-+}\otimes \mathbf{1}_{-+}
\end{equation*}%
a similar conclusion is valid if taking the top-quark in $\mathbf{16}_{++}$.

\  \  \  \

\emph{Breaking} \emph{symmetries in} $SO_{10}\times \mathbb{D}_{4}$- \emph{%
models}\newline
The descent from $SO_{10}\times \mathbb{D}_{4}$- theory to $SU_{5}\times
\Gamma $-models with discrete symmetries $\Gamma \subset \mathbb{D}_{4}$
follows by using U$\left( 1\right) _{X}$ flux to pierce the matter $\mathbf{%
16}_{{}}$ and Higgs $\mathbf{10}_{{}}$ curves as done in \textrm{\cite{1B}}.
For matter sector for example, we start from $M_{16}$ multiplets $\left \{
\mathbf{16}_{i}\right \} _{M_{16}}$ and use $N$ flux units of U$\left(
1\right) _{X}$ to pierce the curve package; and decompose it like $\left \{
\mathbf{10}\right \} _{M_{16}}\oplus \left \{ \mathbf{5}\right \}
_{M_{16}-N}\oplus \left \{ \mathbf{1}\right \} _{M_{16}+N}$; similar
decompositions are valid for the Higgs sector; explicit MSSM- like models
using discrete group character approach will be given in \textrm{\cite{2C}}.
One can also break the $\mathbb{D}_{4}$ monodromy group down to $\mathbb{Z}%
_{4}$ subgroup by using flux and following the same character based method
described in this paper.

\begin{acknowledgement}
Saidi would like to thank the ICTP- Senior Associate programme for
supporting his stay at ICTP, the International Centre for Theoretical
Physics, Trieste Italy; where this work has been completed.
\end{acknowledgement}


\begin{thebibliography}{99}
\bibitem{1A} C. Beasley, J. J. Heckman and C. Vafa, \emph{GUTs and
Exceptional Branes in F-theory - I}, JHEP 0901 (2009) 058 [arXiv:0802.3391
[hep-th]].

\bibitem{2A} R. Donagi and M. Wijnholt, \emph{Model Building with F-Theory},
Adv. Theor. Math.Phys. 15 (2011) 1237 [arXiv:0802.2969 [hep-th]]

\bibitem{3A} C. Beasley, J. J. Heckman and C. Vafa, \emph{GUTs and
Exceptional Branes in F-theory - II}: Experimental Predictions," JHEP 0901
(2009) 059 [arXiv:0806.0102 [hep-th]],

\bibitem{4A} R. Donagi and M. Wijnholt, \emph{Breaking GUT Groups in F-Theory%
}, Adv. Theor.Math. Phys. 15 (2011) 1523 [arXiv:0808.2223 [hep-th]].

\bibitem{5A} R. Donagi and M. Wijnholt, \emph{Higgs Bundles and UV
Completion in F-Theory}, Commun. Math. Phys. 326 (2014) 287 [arXiv:0904.1218
[hep-th]],

\bibitem{6A} J. Marsano, N. Saulina and S. Schafer-Nameki, \emph{%
Monodromies, Fluxes, and Compact Three-Generation F-theory GUTs,}\ JHEP
\textbf{0908} (2009) 046 [arXiv:0906.4672 [hep-th]].

\bibitem{7A} J.J. Heckman, A. Tavanfar and C. Vafa, \emph{The Point of E(8)
in F-theory GUTs},\ JHEP \textbf{1008} (2010) 040 [arXiv:0906.0581 [hep-th]].

\bibitem{8A} A. Maharana and E. Palti, \emph{Models of Particle Physics from
Type IIB String Theory and F-theory: A Review},\ Int.\ J.\ Mod.\ Phys.\ A
\textbf{28} (2013) 1330005 [arXiv:1212.0555 [hep-th]],

\bibitem{1B} I. Antoniadis and G. K. Leontaris, \emph{Building SO(10) models
from F-theory},\ JHEP \textbf{1208} (2012) 001 [arXiv:1205.6930 [hep-th]],

\bibitem{2B} Athanasios Karozas, Stephen F. King, George K. Leontaris,
Andrew K. Meadowcroft, \textquotedblleft Phenomenological implications of a
minimal F-theory GUT with discrete symmetry,\textquotedblright \
arXiv1505:009337 [hep-ph]],

\bibitem{3B} Athanasios Karozas, Stephen F. King, George K. Leontaris and
Andrew Meadowcroft, \emph{Discrete Family Symmetry from F-Theory GUTs},\
[arXiv:1406.6290 [hep-ph]],

\bibitem{4B} I. Antoniadis and G.K. Leontaris, \emph{Neutrino mass textures
from F-theory},\ Eur.\ Phys.\ J.\ C \textbf{73} (2013) 2670 [arXiv:1308.1581
[hep-th]],

\bibitem{5B} Sergio Cecotti, Miranda C.N. Cheng, Jonathan J. Heckman, Cumrun
Vafa\textrm{,} \emph{Yukawa Couplings in F-theory and Non-Commutative
Geometry,} arXiv:0910.0477,

\bibitem{6B} Patrick Morandi, \textquotedblleft Field and Galois
Theory\textquotedblright , Springer,1996,

\bibitem{8B} H. Ishimori, T. Kobayashi, H. Ohki, H. Okada, Y. Shimizu and M.
Tanimoto,\textquotedblleft Non-Abelian Discrete Symmetries in Particle
Physicists,\textquotedblright \ [arXiv:1003.3552 [hep-th]]

\bibitem{1C} R. Ahl Laamara, L.B Drissi, F.Z Hassani, E.H Saidi, A.A
Soumail, Nucl.Phys.B847:275-296,2011, arXiv:1011.3299,\newline
L.B Drissi, F.Z Hassani, H. Jehjouh, E.H Saidi, Phys.Rev.D81:105030,2010,
arXiv:1008.2689,

\bibitem{2C} R. Ahl Laamara, M. Miskaoui, E.H Saidi, \emph{MSSM-like from SU}%
$_{5}\times \mathbb{D}_{4}$ \emph{Model}, LPHE-MS preprint (2015),

\bibitem{1Y} Wiliam Fulton, Joe Harris, \emph{Young Tabeaux with
Applications to Representation Theory and Geometry}, Springer-Verlag (1991),

\bibitem{1X} H. Hayashi, T. Kawano, Y. Tsuchiya and T. Watari, \emph{Flavor
Structure in F-theory Compactifications}, JHEP 1008, 036 (2010)
[arXiv:0910.2762 [hep-th]].
\end{thebibliography}
\end{document}